\documentclass[aps,pra,letterpaper,twocolumn,showpacs,superscriptaddress,floatfix,longbibliography]{revtex4-1}

\usepackage[english]{babel}
\usepackage{amsmath}
\usepackage{color}
\usepackage{epsfig}
\usepackage{graphicx}
\usepackage{bm}
\usepackage{times,amsmath,amssymb}
\usepackage{subfigure}
\usepackage{amsfonts}
\usepackage{amssymb}
\usepackage{color}

\usepackage{mathtools}
\usepackage{empheq}
\usepackage{dsfont}

\newcommand{\tr}{\mathop{\mathrm{tr}}\limits}
\newcommand{\trb}{\mathop{\mathrm{tr}_{1,N}}\limits}
\newcommand{\trzero}{\mathop{\mathrm{tr}_{\mathcal{H}_0}}\limits}
\newcommand{\truno}{\mathop{\mathrm{tr}_{\mathcal{H}_1}}\limits}

\definecolor{myred}{RGB}{168,5,14}

\allowdisplaybreaks

\global\long\def\bra#1{\langle #1 |}
\global\long\def\ket#1{| #1 \rangle }
\global\long\def\braket#1#2{\left\langle #1|#2\right\rangle }

\global\long\def\al{\alpha} \global\long\def\be{\beta}
 \global\long\def\de{\delta}
 
\global\long\def\th{\theta}

\global\long\def\si{\sigma} \global\long\def\vfi{\varphi}

\newcommand{\traccazero}{\mathop{\mathrm{tr}_{\mathcal{H}_0}}\limits}

\global\long\def\bege{\begin{equation}}
  \global\long\def\ende{\end{equation}}

\global\long\def\begal{\begin{align}}
  \global\long\def\endal{\end{align}}

\begin{document}

\title{Effective quantum Zeno dynamics in dissipative quantum systems
}

\author{Vladislav Popkov} \address{ HISKP, University of Bonn,
  Nussallee 14-16, 53115 Bonn, Germany.} \address{ Institut f\"{u}r
  Teoretische Physik, Universit\"{a}t zu K\"{o}ln, Z\"ulpicher
  str. 77, K\"{o}ln, Germany.}  \address{ Centro Interdipartimentale
  per lo studio di Dinamiche Complesse, Universit\`a di Firenze, via
  G.  Sansone 1, 50019 Sesto Fiorentino, Italy } \author{Simon Essink}
\address{ HISKP, University of Bonn, Nussallee 14-16, 53115 Bonn,
  Germany.}  \author{Carlo Presilla} \address{Dipartimento di Fisica,
  Sapienza Universit\`a di Roma, Piazzale Aldo Moro 2, Roma 00185,
  Italy} \address{Istituto Nazionale di Fisica Nucleare, Sezione di
  Roma 1, Roma 00185, Italy} \vspace{10pt} \author{Gunter Sch\"utz}
\address{ICS, Forschungszentrum J\"uelich GmbH Wilhelm-Johnen-Stra\ss
  e 52428 J\"uelich, Germany }

\begin{abstract}
  We investigate the time evolution of an open quantum system
  described by a Lindblad master equation with dissipation acting only
  on a part of the degrees of freedom ${\cal H}_0$ of the system, and
  targeting a unique dark state in ${\cal H}_0$.  We show that, in the
  Zeno limit of large dissipation, the density matrix of the system
  traced over the dissipative subspace ${\cal H}_0$, evolves according
  to another Lindblad dynamics, with renormalized effective
  Hamiltonian and weak effective dissipation.  This behavior is
  explicitly checked in the case of Heisenberg spin chains with one or
  both boundary spins strongly coupled to a magnetic reservoir.
  Moreover, the populations of the eigenstates of the renormalized
  effective Hamiltonian evolve in time according to a classical Markov
  dynamics. As a direct application of this result, we propose a
  computationally-efficient exact method to evaluate the
  nonequilibrium steady state of a general system in the limit of
  strong dissipation.
\end{abstract}

\maketitle

\section{Introduction}
A quantum system interacting with an environment is, under a Markov
assumption, well described by a Lindblad master equation
(LME)~\cite{Lindblad,GKS}.  It follows that the reduced density matrix
(RDM) of the system undergoes a coherent and dissipative
evolution~\cite{Petruccione,ClarkPriorMPA2010}.  If the coherent and
dissipative parts of LME do not depend on time, then, after a
transient, the system reaches a (unique) nonequilibrium steady state
(NESS), which is independent of the initial conditions.  Even if the
NESS is trivial, the relaxation dynamics may not be: specially if a
large dissipation-free subspace exists, the NESS can be approached
through a complicated multi-stage evolution.

If the dissipation time-scales are short in comparison with the
coherent evolution times, then the so-called quantum Zeno regime
occurs.  Quantum Zeno
effect~\cite{Misra1977,ZenoStaticsExperimentalReview} predicts an
inhibition of quantum transitions in a quantum system subjected to
frequent measurements. It has been observed experimentally, in various
setups \cite{Itano1990,Kwiat1995,Signoles2014,Schafer2014,Patil2015}.
Applications of Zeno effect include dissipation-protected realization
of quantum gates~\cite{Beige2000}, engineering of nontrivial quantum
states and implementation of universal quantum
computations~\cite{Verstraete2009,Yi2012,Elliott2015,Winkler2006}
creating quantum simulators~\cite{Stannigel2013}, localization of a
single atom in a lattice~\cite{Ashida2015}, realization of exotic
effective dynamics~\cite{Lee2014,Elliott2016a}, development of
theoretical tools for a real-time observation of quantum many-body
dynamics~\cite{Ashida2017}.

It is well recognized that the evolution of a system near the Zeno
limit is not frozen but can proceed via Raman-like processes involving
virtual levels, which couple states within a given Zeno
subspace~\cite{Kozlowski2016,Elliott2016b}, while the occupation of
the virtual levels remains negligible.

In more details, one can distingish three stages of relaxation,
occurring at different time scales.  On the shortest time scale, only
the degrees of freedom directly affected by the dissipation, relax to
their stationary values. On the second, intermediate time scale, an
effective coherent evolution takes place, governed by a
dissipation-projected Hamiltonian~\cite{2014Venuti}. Finally, on the
longest time scale, all system characteristics relax to their
stationary values.

In this paper, we focus on the third stage of evolution and derive an
effective dynamics of the system in the decoherence-free subspace.  It
happens that, in the assumed Zeno regime, and under the non-degeneracy
assumption for the local kernel of the dissipator (\ref{TargetZeno}),
this dynamics is also of Lindblad type.  As an application, we
demonstrate that the spectrum of the reduced density matrix, which
does not change on the intermediate time scale, on the longest time
scale evolves according to a classical Markov process, with generator
$F$ computable from the LME entries.

\section{Main results}
Consider an open quantum system, with finite Hilbert space ${\cal H}$,
under strong dissipation acting only on a subspace $\mathcal{H}_0$ of
the degrees of freedom, described by the Lindblad master equation,
\begin{align}
  \frac{\partial \rho(\tau) }{\partial \tau} = -\frac{i}{\hbar}
  \left[H,\rho(\tau)\right] + \Gamma {\cal D} [\rho(\tau)].
  \label{LME}
\end{align}
Let the dissipation-free subspace be ${\cal H}_1$, ${\cal H}={\cal
  H}_0\otimes {\cal H}_1$, and denote by $\trzero$ and $\truno$ the
trace over $\mathcal{H}_0$ and $\mathcal{H}_1$, having dimensions
$d_0$ and $d_1$, respectively.  We assume the Lindblad dissipator
$\mathcal{D}$ to target a unique state $\psi_0\in\mathcal{H}_0$,
namely,
\begin{align}
  \left(\truno \mathcal{D}\right) \psi_0 = 0.
  \label{TargetZeno}
\end{align}
The aim of this paper is to show that, in the Zeno limit, when
the effective dissipation strength $\Gamma$ is much stronger than the
unitary part of the evolution, the solution of the problem (\ref{LME})
for all times $\tau>O(1)$ has the approximate form
\begin{align}
  \rho(\tau) =\psi_0\otimes R(\tau),
  \nonumber
\end{align}
where $R(\tau)=\trzero \rho(\tau)$ evolves according to another LME
\begin{align}
  \frac{\partial R(\tau)}{\partial \tau} = - \frac{i}{\hbar} \left[
    \tilde{H},R(\tau) \right] + \frac{1}{\Gamma} \tilde{\mathcal{D}}
  [R(\tau)].
  \label{LMEforR}
\end{align}
More precisely, we demonstrate that
\begin{align}
  \| \rho(\tau) - \psi_0\otimes R(\tau) \| = O\left(
    \frac{1}{\Gamma}\right),
  \label{RESrho(t)}
\end{align}
for $\Gamma\to\infty$ and for all times $\tau \gg 1/\Gamma$.  The
choice of the norm $\| \cdot \|$ is rather arbitrary.  Note that the
LMEs (\ref{LME}) and (\ref{LMEforR}), besides being defined in terms
of different Hamiltonians and dissipators, have dissipation strength
$\Gamma$ and $1/\Gamma$, respectively.

Using $1/\Gamma\ll 1$ as a small parameter, we obtain the above result
by writing the Dyson series for the Liouvillian dynamics associated to
the LME (\ref{LME}).  We start rescaling the time $\Gamma \tau = t$ in
the original LME. In the limit of strong dissipation $\Gamma \gg 1$,
we obtain an equation with a perturbative term,
\begin{equation}
  \label{LMErescaled}
  \frac{\partial \rho}{\partial t} =
  \mathcal{D}[\rho] -\frac{i}{\Gamma} \left[ H,\rho \right]
  = (\mathcal{L}_0 + K)\rho = \mathcal{L} \rho,
\end{equation}
where $\mathcal{L}=\mathcal{L}_0+K$ and the linear operators
$\mathcal{L}_0$ and $K = -(i/\Gamma) \left[ H, \cdot \right]$ denote
the dissipator and the commutator, respectively. The formal solution
of Eq.~(\ref{LMErescaled}) is
\begin{equation}
  \rho(t) = e^{\mathcal{L} t} \rho(0) = \mathcal{E}(t) \rho(0),
  \label{formal_sol_rho}
\end{equation}
where the propagator $\mathcal{E}(t)$ satisfies
\begin{equation}
 \mathcal{E}(t) = e^{\mathcal{L}_0 t} \left( 1+
 \int_0^t dt_1 e^{-\mathcal{L}_0 t_1} K \mathcal{E}(t_1)\right).
 \label{expansion_semigroup}
\end{equation}
Iterating Eq.~(\ref{expansion_semigroup}) we get the Dyson
expansion. Up to the second order we obtain
\begin{align}
  \mathcal{E}(t) =&\ e^{\mathcal{L}_0 t} \left(1 + \int_0^t \!\!\!dt_1
    e^{-\mathcal{L}_0 t_1} K e^{\mathcal{L}_0 t_1} \right.
  \nonumber\\ &+ \left.  \int_0^t \!\!\!dt_1 e^{-\mathcal{L}_0 t_1} K
    e^{\mathcal{L}_0 t_1} \int_0^{t_1} \!\!\!dt_2 e^{-\mathcal{L}_0
      t_2} K e^{\mathcal{L}_0 t_2} + \dots \right).
\end{align}
Introduce the spectral projection $\mathcal{P}_0$ onto the kernel of
the dissipator $\mathcal{L}_0$, namely,
$\mathcal{P}_0=\lim_{t\rightarrow \infty} \exp(\mathcal{L}_0 t)$.
Define also its complement $\mathcal{Q}_0=
I_\mathcal{H}-\mathcal{P}_0$, where $I_\mathcal{H}$ is the identity
operator in the space $\mathcal{H}$. Obviously, $\mathcal{P}_0
\mathcal{Q}_0 = 0$.  If $1/\Gamma$ is small, the dissipative part of
the Lindbladian constrains the system to a decoherence-free
subspace. In fact, the leakage outside of decoherence-free subspace
(defined as the subspace belonging to the dissipator Kernel) can be
rigorously proven to be negligible, see Ref.~\cite{2014Venuti}. Therefore,
we shall only consider an evolution inside the decoherence-free
subspace, which is given by the propagator $\mathcal{P}_0
\mathcal{E}(t) \mathcal{P}_0$.  Performing the
calculations as indicated in Appendix~\ref{SuppMat}, we obtain
\begin{align}
  \mathcal{P}_0\mathcal{E}(t) \mathcal{P}_0 =&\ \mathcal{P}_0+ t
  \mathcal{P}_0 K \mathcal{P}_0 + \frac{t^2}{2!}(\mathcal{P}_0 K
  \mathcal{P}_0)^2 \nonumber \\ &- t \mathcal{P}_0 K\mathcal{Q}_0
  \mathcal{S} K \mathcal{P}_0 + \ldots,
  \label{Dyson2-Final}
\end{align}
where $\ldots$ is the contribution from the remaining orders of the
Dyson expansion, and $\mathcal{S}$ is the pseudo-inverse of the
dissipator,
\begin{align}
  \mathcal{L}_0 \mathcal{S} = \mathcal{S} \mathcal{L}_0 =
  \mathcal{Q}_0.
\end{align}
Note that the first three terms in Eq.~(\ref{Dyson2-Final}) can be
exponentiated, as $\mathcal{P}_0 \exp(t \mathcal{P}_0 K
\mathcal{P}_0)$. They all describe a unitary dynamics within the
decoherence-free subspace, as is seen by applying the propagator
$\mathcal{P}_0 K \mathcal{P}_0$ on a state $\rho=\psi_0 \otimes R$,
\begin{align}
  \mathcal{P}_0 K \mathcal{P}_0 \rho &= - \frac{i}{\Gamma}
  \mathcal{P}_0[H,\psi_0 \otimes R] \nonumber \\ &= - \frac{i}{\Gamma}
  \psi_0 \otimes \trzero \left( H(\psi_0 \otimes R) - (\psi_0 \otimes
    R)H \right) \nonumber \\ &= -\frac{i}{\Gamma} \psi_0 \otimes
  \left[ h_D, R \right],
\end{align}
where $ h_D$ is the dissipation-projected Hamiltonian
\begin{align}
  h_D = \trzero \left( \left(\psi_0 \otimes I_{\mathcal{H}_1}\right)H
  \right).
  \label{h_D}
\end{align}
Since the operator $K$ is proportional to the small
parameter $1/\Gamma$, we conclude that the terms $t \mathcal{P}_0 K
\mathcal{P}_0$ and $\frac{t^2}{2!}(\mathcal{P}_0 K \mathcal{P}_0)^2$
give a contribution $O(1)$ to the propagator for times $t \sim
O(\Gamma)$, while the last term $ - t \mathcal{P}_0 K\mathcal{Q}_0
\mathcal{S} K \mathcal{P}_0$ contributes $O(1)$ changes to the
propagator for $t \sim O(\Gamma^2)$.  The physical interpretation of
Eq.~(\ref{Dyson2-Final}) is thus as follows.  One observes
three different processes, taking place at different time
scales $\tau=t/\Gamma$: (i) at short times $\tau \sim 1/\Gamma$, the
system is projected onto the decoherence-free subspace; (ii) at
intermediate times $\tau \sim 1$, the evolution inside the
decoherence-free subspace is unitary $\mathcal{P}_0 K \mathcal{P}_0
\sim -i \psi_0 \otimes [h_{D},\cdot]$; (iii) at large times $\tau \sim
\Gamma$ the term $t\mathcal{P}_0 K\mathcal{Q}_0 \mathcal{S} K
\mathcal{P}_0$ sets in.  Note that the slowest part of the evolution,
taking place at the longest time scale, cannot by any means be ignored
since it is the only part containing a relaxation towards the NESS.
In fact, the unitary evolution alone governed by the effective
Hamiltonian (\ref{h_D}), does not lead to any relaxation.

\begin{figure}[t]
  \begin{center}
    \includegraphics[width=0.99\columnwidth]{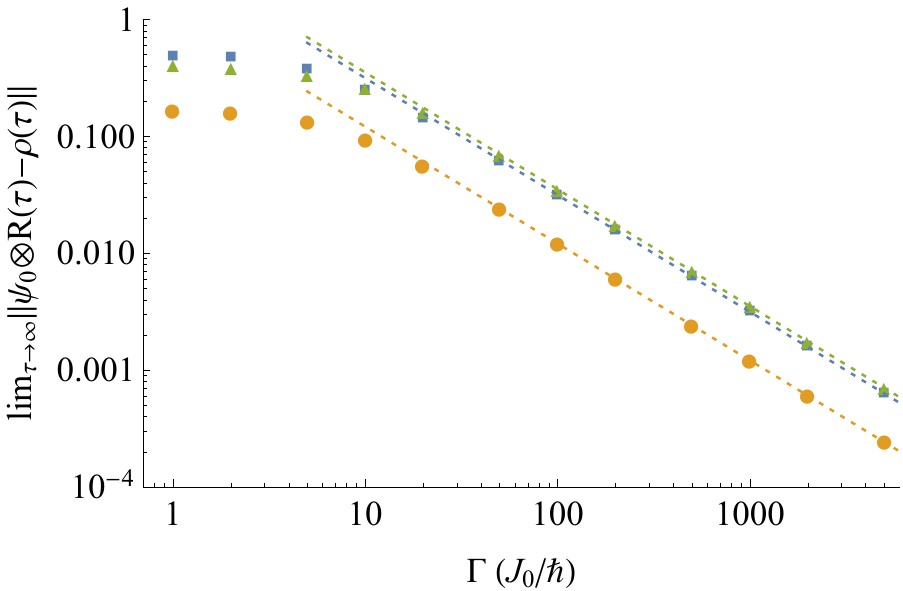}
  \end{center}
  \caption{Asymptotic error (Euclidean norm)
    $\lim_{\tau\to\infty}\|\psi_0\otimes R(\tau)-\rho(\tau)\|$ as a
    function of the dissipation strength $\Gamma$ for the $XYZ$
    Heisenberg spin chain with dissipation acting on the first and
    last spins.  Here, $R(\tau)$ is the solution of
    Eq.~(\ref{LMEeffective}) whereas $\rho(\tau)$ is the solution of
    Eq.~(\ref{LME}).  Parameters: $N=4$, $J_x=J_0,J_y=2.2 J_0,J_z=0.77
    J_0$ for all data-points.  Triangles:
    $\th_L=\vfi_L=\th_R=\vfi_R=0$, $\mu_L=0.9, \mu_R=0.7$.  Squares:
    $\th_L=\pi/3$,$\vfi_L=\pi/4$, $\th_R=3\pi/7$, $\vfi_R=4\pi/15$,
    $\mu_L=0.9, \mu_R=0.7$. Circles: $\th_L=\pi/3$,$\vfi_L=\pi/4$,
    $\th_R=3\pi/7$, $\vfi_R=4\pi/15$, $\mu_L=0.5, \mu_R=-0.3$. The
    straight lines with slope $1/\Gamma$ are guides to the eye. }
  \label{fig1}
\end{figure}

To derive the evolution equation from the Dyson expansion, assume the
system to start in the dissipation-free subspace, i.e., $\rho(0) =
\mathcal{P}_0 \rho(0)$. This is equivalent to assuming the factorized
initial state $\rho(0)=\psi_0 \otimes R(0)$.  The time evolution
inside the decoherence-free subspace is given by $\mathcal{P}_0
\mathcal{E}(t) \mathcal{P}_0 [\psi_0 \otimes R(0)]= \psi_0 \otimes
R(t)$.  We obtain the evolution equation in differential form
considering $\lim_{t\rightarrow 0} [\rho(t)-\rho(0)]/t= \partial
\rho/\partial t$.  Using the Dyson expansion, tracing over
$\mathcal{H}_0$, and rescaling the time $t/\Gamma=\tau$, we obtain
\begin{align}
  \frac {\partial R}{\partial \tau} &= -i \left[h_D,R(\tau) \right]+
  \frac{1}{\Gamma}W
  \label{LMEeffective}\\
  W &= -\Gamma^2 \trzero \left(\mathcal{P}_0 K \mathcal{Q}_0 S K
    \mathcal{P}_0 \rho \right)
  \label{DefW}
\end{align}
Equation~(\ref{LMEeffective}) is valid for time scales beyond the
shortest one, i.e., $\tau \gg 1/\Gamma$. The total error of the
effective description (\ref{LMEeffective}) of the evolution
$\rho(0)\rightarrow \rho(\tau)\approx \psi_0\otimes R(\tau)$ for large
$\Gamma$ results from two contributions: a leakage outside the
dissipation-free subspace and higher order dissipation terms, both
contributions being generically of order $1/\Gamma$, see also
Fig.~\ref{fig1}.

To evaluate $W$ from Eq.~(\ref{DefW}), we make two assumptions: (i)
the kernel of $\mathcal{L}_0$ is one-dimensional, i.e., the eigenvalue
$0$ of the dissipator is non-degenerate,
\begin{align}
  \mathcal{L}_0 \psi_0&=0;
  \label{Dyson2-0}
\end{align}
(ii) $\mathcal{L}_0$ is diagonalizable, i.e., a basis $\psi_k$ (not
necessarily orthogonal) exists,
\begin{align}
  \mathcal{L}_0 \psi_k&=\xi_k \psi_k.
  \label{Dyson2-1}
\end{align}
Note that $\psi_k^\dagger$ are also eigenvectors of the dissipator,
with eigenvalues $\xi_k^*$, namely, $\mathcal{L}_0
\psi_k^\dagger=\xi_k^* \psi_k^\dagger$.  We also introduce a
complementary basis $\vfi_k$, trace-orthonormal to the basis $\psi_j$,
\begin{align}
  \tr(\vfi_k \psi_j)&=\delta_{k,j}.
  \label{Dyson2-orthonormalBasis}
\end{align}
Hereafter, we work in the representation in which $\psi_k,\vfi_k$ are
square matrices.

First, we note that the action of $\mathcal{P}_0$ on the arbitrary
element $X \in \mathcal{H}$ is
\begin{align}
  \mathcal{P}_0 X &=\psi_0 \otimes \trzero X.
  \label{Dyson2-P0action}
\end{align}
In fact, due to the definition of $\mathcal{P}_0$ we have
\begin{align}
  \mathcal{P}_0 X &=\lim_{t\rightarrow \infty} e^{\mathcal{L}_0 t} X
  =\lim_{t\rightarrow \infty} e^{\mathcal{L}_0 t} \sum_{k} \psi_k
  \otimes x_k \nonumber \\ &= \sum_{k} \lim_{t\rightarrow \infty}
  e^{\xi_k t} \psi_k \otimes x_k = \psi_0 \otimes x_0,
  \label{Dyson2-P0action1}
\end{align}
since the real part of all $\xi_k$ for $k>0$ is strictly negative.  In
the decomposition $X=\sum_{k} \psi_k \otimes x_k$, the element $x_0$
can be found using the trace-orthonormal basis $\vfi_k$ as
$x_0=\trzero(\vfi_0\otimes I_{\mathcal{H}_1}) X$. The element $\vfi_0$
of this basis, satisfying $\tr (\vfi_0 \psi_k)=\de_{k,0}$, can always
be chosen as the unit matrix, $\vfi_0=I_{\mathcal{H}_0}$, since all
the eigenfunctions of the dissipator with nonzero eigenvalues are
traceless, and $\tr \psi_0=1$.  Substituting $x_0= \trzero X$ in
Eq.~(\ref{Dyson2-P0action1}), we obtain Eq.~(\ref{Dyson2-P0action}).

It is convenient to define the Hamiltonian decomposition
\begin{align}
  H &= \sum_n \vfi_n \otimes g_n=\sum_n \vfi_n^\dagger \otimes
  g_n^\dagger,
  \label{Dyson2-Decomposition}\\
  g_k &= \trzero ((\psi_k \otimes I_{\mathcal{H}_1})
  H). \label{Dyson2-hk}
\end{align}
We have, step by step,
\begin{align*}
  \mathcal{P}_0 \rho(0)&=  \rho(0),\\
  (\Gamma K) \mathcal{P}_0 \rho(0) &= -i\left[ H, \rho(0) \right]
  \nonumber \\ &= -i \sum_{m,n} \left(
    C_{mn} \psi_m^\dagger \otimes (g_n R) - \mathrm{H.c.} \right),\\
  \mathcal{Q}_0 S (\Gamma K) \mathcal{P}_0 \rho(0) &= -i \!\!\!
  \sum_{m>0,n} \frac{1}{\xi_m^*} \left( C_{mn} \psi_m^\dagger \otimes
    (g_n R) - \mathrm{H.c.} \right),
\end{align*}
where
\begin{align}
  C_{mn}&= \tr \left( \vfi_m^\dagger \vfi_n \psi_0 \right).
  \label{Dyson2-CoeffCmn}
\end{align}
Since $\vfi_0=I_{\mathcal{H}_0}$, the coefficients $C_{mn}$ satisfy
\begin{align}
  C_{0n} &= C_{n0}=\de_{0,n}.
  \label{Dyson2-C0n}
\end{align}
In the last step, using Eqs.~(\ref{Dyson2-orthonormalBasis}) and
(\ref{Dyson2-P0action}), we arrive at
\begin{align}
  W &= \!\!\!\! \sum_{m>0,n>0} \left( \frac{C_{mn}}{-\xi_m^*} \left(
      -g_m^\dagger g_n R + g_n R g_m^\dagger \right) + \mathrm{H.c.}
  \right).
  \label{Dyson2-6}
\end{align}
Note that the term $n=0$ does not appear in the sum (\ref{Dyson2-6})
because of Eq.~(\ref{Dyson2-C0n}).  Using the substitution
$-C_{mn}/\xi_m^*=Y_{mn}=A_{mn}/2 +i B_{mn}$ with
$A_{mn}=Y_{mn}+Y_{nm}^*$ positive matrix and
$B_{mn}=(Y_{mn}-Y_{nm}^*)/(2i)$ Hermitian matrix, and changing the
order of summation in the H.c. term in (\ref{Dyson2-6}), we can put
$W$ in the general Lindbladian form,
\begin{align}
  &W = - i [\tilde{H}_a,R] + \tilde{\mathcal{D}} R,
  \label{Dyson2-W}
  \\
  &\tilde{H}_a = \!\!\!\!\!\! \sum_{m>0,n>0} \!\!\!\!\! B_{mn}
  g_m^\dagger g_n,
  \label{Dyson2-W.H}
  \\
  &\tilde{\mathcal{D}} R = \!\!\!\!\!\! \sum_{m>0,n>0} \!\!\!\!\!
  A_{mn} \left( g_n R g_m^\dagger - \frac{1}{2} g_m^\dagger g_n R-
    \frac{1}{2} R g_m^\dagger g_n \right).
  \label{Dyson2-W.D}
\end{align}

According to Eq.~(\ref{LMEeffective}), from the above expression of
$W$ we conclude that the effective time evolution of the system in the
dissipation-free subspace has the standard Lindblad form of
Eq.~(\ref{LMEforR}), with $\tilde{H}=h_D+\tilde{H}_a/\Gamma$ and the
dissipator $\tilde{\mathcal{D}}/\Gamma$ with $\tilde{\mathcal{D}}$
given by Eq.~(\ref{Dyson2-W.D}). Note that the stronger is
the dissipation $\Gamma$ in the original system, the weaker
is the effective dissipation (of order $1/\Gamma$) in the effective
dynamics~\cite{Carollo2006}.

\section{Heisenberg spin chain with the first spin in a target state}
To illustrate our findings, we consider a system of interacting spins,
with one spin strongly dissipatively coupled to an environment which
targets an arbitrary mixed state $\psi_0$ of that spin. In the
Lindblad formalism, this is achieved via the application of two
Lindblad operators~\cite{ProsenExact2011},
\begin{align}
  L_1 = \sqrt{\frac{1+\mu}{2}} \ket{0^\perp}\bra{0}, \qquad L_2 =
  \sqrt{\frac{1-\mu}{2}} \ket{0}\bra{0^\perp}
  \label{TargetingOneSpin-L1L2},
\end{align}
where $\ket{0}$ is an arbitrary normalized state in
$\mathcal{H}_0\equiv\mathbb{C}_2$, $\braket{0^\perp}{0}=0$ and $\mu$
real parameter with $0 \leq \mu \leq 1$.  The resulting dissipator
$\mathcal{L}_0=\mathcal{D}_{L_1} + \mathcal{D}_{L_2}$, where
\begin{align}
  \mathcal{D}_L X &= L X L^\dagger - \frac{1}{2} (L^\dagger L X + X
  L^\dagger L),
  \label{TargetingOneSpin-Diss}
\end{align}
targets the arbitrary mixed state of a single spin
\begin{align}
  \psi_0 = \frac{1+ \mu}{2} \ket{0}\bra{0} + \frac{1- \mu}{2}
  \ket{0^\perp} \bra{0^\perp}.
  \label{TargetingOneSpin-psi0}
\end{align}
In fact, $\psi_0$ is an eigenvector of the dissipator $\mathcal{L}_0$
with eigenvalue $\xi_0=0$, namely, $\mathcal{L}_0 \psi_0 = 0$. The
other eigenvectors and the corresponding eigenvalues of
$\mathcal{L}_0$ are
\begin{align}
  \psi_1&= \ket{0} \bra{0^\perp}, \qquad  \xi_1=-\frac{1}{2},\\
  \psi_2&= \ket{0^\perp} \bra{0}, \qquad \xi_2=-\frac{1}{2},\\
  \psi_3&= \ket{0} \bra{0}-\ket{0^\perp} \bra{0^\perp}, \qquad
  \xi_3=-1.
  \label{psik}
\end{align}
The trace-orthonormal basis $\vfi_k$ satisfying $\tr(\vfi_k
\psi_m)=\delta_{k,m}$ is given by
\begin{align}
  \vfi_0&= I_{\mathbb{C}_2},\\
  \vfi_1&= \ket{0^\perp} \bra{0},\\
  \vfi_2&= \ket{0} \bra{0^\perp},\\
  \vfi_3&= \frac{1- \mu}{2} \ket{0} \bra{0} - \frac{1+
    \mu}{2}\ket{0^\perp} \bra{0^\perp}.
  \label{TargetingOneSpin-TraceOrthonorm}
\end{align}
Given the explicit form of $\vfi_k,\psi_k$, we readily compute the
coefficients $C_{mn}$ from Eq.~(\ref{Dyson2-CoeffCmn}). The only
nonzero coefficients $C_{mn}$ are the diagonal ones: $C_{00}=1$,
$C_{11}=(1+\mu)/2$, $C_{22}=(1-\mu)/2$, $C_{33}=(1-\mu^2)/4$.
Substituting them into Eq.~(\ref{Dyson2-6}) and using
Eq.~(\ref{Dyson2-W}), we obtain $\tilde{H}_a=0$ and
\begin{align}
  \tilde{\mathcal{D}} &= 2(1+\mu) \mathcal{D}_{g_1} + 2(1-\mu)
  \mathcal{D}_{g_1^\dagger} + \frac{1}{2}(1-\mu^2) \mathcal{D}_{g_3}.
  \label{Mixed-6}
\end{align}
The operators $g_k$, given by Eq.~(\ref{Dyson2-hk}), can be evaluated
afterward the Hamiltonian $H$ of the system is specified.

For definiteness, we consider the coherent part of the dynamics to be
given by an open anisotropic $XYZ$ Heisenberg spin chain, with
Hamiltonian
\begin{align}
  H =\sum_{n=1}^{N-1} \vec{\sigma}_n \cdot (J \vec{\sigma}_{n+1}),
  \label {Applications-Ham}
\end{align}
where $\vec{\si}_{n}=(\si_n^x,\si_n^y,\si_n^z)$ and $J= \mathrm{diag}
(J_x,J_y,J_z)$ is the anisotropy tensor of the exchange interaction.
We parametrize the state $\ket{0}$ via spherical coordinates
$\theta,\varphi$,
\begin{align}
  \ket{0}=\left(
    \begin{array}{c}
      \cos(\th/2) e^{-i \vfi/2}
      \\
      \sin(\th/2) e^{i \vfi/2}
    \end{array}
  \right).
\end{align}
Introducing a standard unit vector in polar coordinates,
\begin{align*}
  \vec{n}(\th,\vfi) = (\sin \th \cos \vfi, \sin \th \sin \vfi, \cos
  \th ),
\end{align*}
and other two unit vectors defined as $\vec{n}' = \vec{n} \left(
  \frac{\pi}{2}-\th,\vfi+\pi \right)$, $\vec{n}'' = \vec{n} \left(
  \frac{\pi}{2},\vfi+\frac{\pi}{2} \right)$, in such a way that the
triplet $\vec{n},\vec{n}',\vec{n}''$ forms an orthonormal basis in the
three-dimensional space, we find
\begin{align}
  g_{1}&= (J\vec{n}') \cdot \vec{\sigma}_1 - i (J\vec{n}'') \cdot
  \vec{\sigma}_1,
  \label{Defg1}\\
  g_{3}&= 2 (J\vec{n}) \cdot \vec{\sigma}_1.
  \label{Defg3}
\end{align}
Note that, after tracing over the spin space of the first site as
indicated in (\ref{Dyson2-hk}), in the above expressions we renumerate
the $N-1$ sites not directly affected by the dissipation as
$1,2,\dots, M=N-1$.  With this convention, the dissipation-projected
Hamiltonian is still an anisotropic $XYZ$ Heisenberg Hamiltonian as
$H$ but with $M$ sites and a boundary field
\begin{align}
  h_{D} = \sum_{j=1}^{M-1} \vec{\sigma}_j \cdot (J \vec{\sigma}_{j+1})
  + (J\vec{n}) \cdot \vec{\sigma}_1.
  \label{DefhD}
\end{align}
The Hamiltonian (\ref{DefhD}) and the dissipator defined by
Eq.~(\ref{Mixed-6}) determine the effective LME which governs the time
evolution of the reduced density matrix $R(\tau)$ in the Zeno limit.

\section{Heisenberg spin chain with the first and the last spins in a
  target state}
\label{sec::BoundaryDissipation}

Previous results straightforwardly extend to more general setups.  As
an example, consider the same spin chain discussed above with
dissipation acting only at the boundary spins $1$ and $N$.  Within
this setup, and by tuning of the Hamiltonian parameters, one can
generate, in the Zeno limit, a bulk NESS ranging from a maximally
mixed state~\cite{PopkovSalernoLivi2015} to a pure spin-helix state
carrying ballistic current of magnetization
\cite{2016PopkovPresilla,2017PopkovPresillaSchmidt}.  Here we assume
the dissipation to target generic spin-$1/2$ mixed states, $\psi_0^L$
and $\psi_0^R$, at the sites $1$ and $N$, respectively,
\begin{align}
  \psi_0^{L} &= \frac{1+ \mu_{L}}{2} \ket{0_{L}}\bra{0_{L}} + \frac{1-
    \mu_{L}}{2} \ket{0_{L}^\perp} \bra{0_{L}^\perp},
  \\
  \psi_0^{R} &= \frac{1+ \mu_{R}}{2} \ket{0_{R}}\bra{0_{R}} + \frac{1-
    \mu_{R}}{2} \ket{0_{R}^\perp} \bra{0_{R}^\perp}.
\end{align}
As discussed above, this is realized by applying two Lindblad
operators, of the form (\ref{TargetingOneSpin-L1L2}), at each end of
the chain with parameters $\mu_L$ and $\mu_R$, respectively.

Overall the dissipation targets a state which is the product of the
states targeted at the left and right boundaries, $\psi_0=\psi_0^L
\otimes \psi_0^R $. The eigenvalues of the full dissipator are the sum
of the eigenvalues of the left and right boundary dissipators
separately, $\xi_{m_L}+\xi_{m_R}$, and the respective eigenvectors are
$\psi_{m_L,m_R}=\psi^L_{m_L}\otimes \psi^R_{m_L}$, where the
individual $\psi^{L,R}_{m}$ have the form (\ref{psik}).  The
Hamiltonian decomposition in terms of the trace-orthonormal basis for
the left and right dissipators, $\vfi^L_{n_L}$,$\vfi^R_{n_R}$, now
reads
\begin{align}
  &H = \sum_{n_L,n_R} \vfi^L_{n_L} \otimes g_{n_L,n_R} \otimes
  \vfi^R_{n_R},
  \label{Dyson2-Decomposition2}
  \\
  &g_{n_L,n_R}= \trb [ ( \psi^L_{n_L} \otimes I^{2^{N-1}} ) H (
  I^{2^{N-1}} \otimes \psi^R_{n_R} ) ].
  \label{hkcomposite}
\end{align}
We can therefore apply the general formula (\ref{Dyson2-6}), with
$\xi_{m} \to \xi_{m_L}+\xi_{m_R}$ and $g_n \to g_{n_L,n_R}$.  Note
that, due to the locality of the interactions, $g_{n_L,n_R}=0$ if $n_L
n_R \neq 0$. After some algebra, and using Eq.~(\ref{Dyson2-C0n}), we
obtain that Eq.~(\ref{Dyson2-6}) splits into the sum of two
contributions, associated to the left and right ends of the chain,
\begin{align}
  W &= \tilde{\mathcal{D}}_L R + \tilde{\mathcal{D}}_R R,
  \label{Applications-2}
\end{align}
where, according to (\ref{Mixed-6}),
\begin{align*}
  \tilde{\mathcal{D}}_L &= 2(1+\mu_L) \mathcal{D}_{g_{10}} +
  2(1-\mu_L) \mathcal{D}_{g_{10}^\dagger} + \frac{1} {2}(1-\mu_L^2)
  \mathcal{D}_{g_{30}},
  \\
  \tilde{\mathcal{D}}_R &= 2(1+\mu_R) \mathcal{D}_{g_{01}} +
  2(1-\mu_R) \mathcal{D}_{g_{01}^\dagger} + \frac{1} {2}(1-\mu_R^2)
  \mathcal{D}_{g_{03}}.
\end{align*}
Also in the present case, $W$ does not have coherent contributions of
the kind (\ref{Dyson2-W.H}).

The operators $g_{k0},g_{0k}$, as well as the dissipation-projected
Hamiltonian $h_{D}$, can be evaluated exactly as in the previous case
of a single spin directly affected by the dissipation. The result is
expressed in terms of the parameters $\mu_L,\mu_R$ and of the polar
coordinates $\th_L,\vfi_L$ and $\th_R,\vfi_R$ which define the states
$\ket{0_L}$ and $\ket{0_R}$.  In particular, the Hamiltonian $h_D$ is
again a $XYZ$ Hamiltonian with $M=N-2$ spins, namely, those not
directly affected by the dissipation, with two boundary terms relative
to the spins 1 and $M$. Explicit formulas will be given elsewhere.  In
Figs.~\ref{fig1}, \ref{fig2} and \ref{fig3} we illustrate the behavior
of the resulting effective LME in comparison with the exact dynamics
of the system.

\section{Evaluation of the NESS in the Zeno limit}
\begin{figure}[t]
  \begin{center}
    \includegraphics[width=0.99\columnwidth]{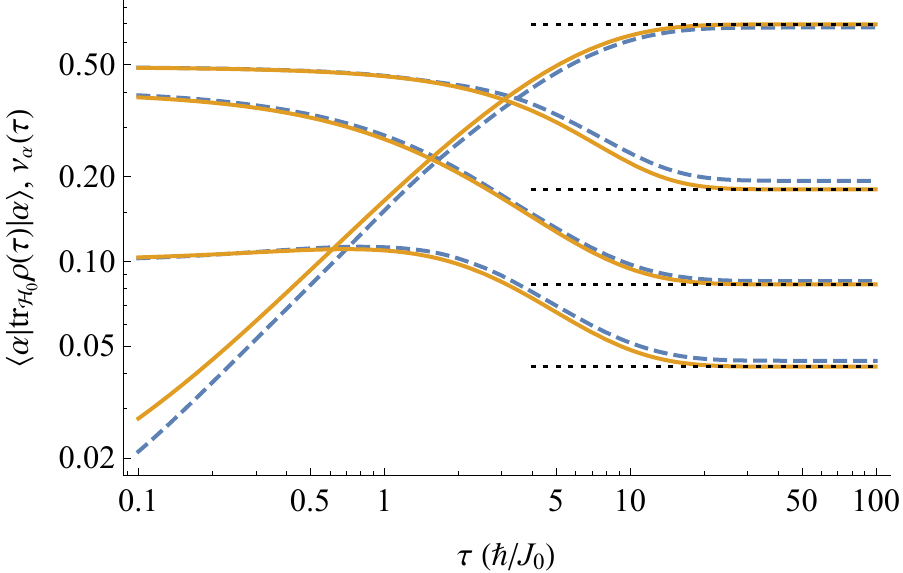}
  \end{center}
  \caption{ Populations of the eigenstates of $h_D$, $\bra{\al}
    \trzero\rho(\tau) \ket{\al} $ (dashed lines) and solutions
    $\nu_\al(\tau)$ of the classical Markov equation
    (\ref{NESS-ClassME}) (solid lines) as a function of time $\tau$
    for the $XYZ$ Heisenberg spin chain with dissipation acting on the
    first and last spins.  We set $\Gamma=50 J_0/\hbar$ and all
    the other parameters are as in Fig.~\ref{fig1}, case of squares.
    The initial condition is $\rho(0)=\psi_0^L \otimes R(0) \otimes
    \psi_0^R$, where $R(0)$ is a diagonal matrix with entries $0.01,
    0.4, 0.1, 0.49$ in the $h_D$ basis.  The straight dotted lines
    indicate the exact eigenvalues of $\trzero\rho(\tau)$ for $\tau
    \rightarrow \infty$ in Zeno limit, computed from the Markov
    process with the rates~(\ref{NESS-ClassME1}).}
  \label{fig2}
\end{figure}

\begin{figure}[t]
  \begin{center}
    \includegraphics[width=0.99\columnwidth]{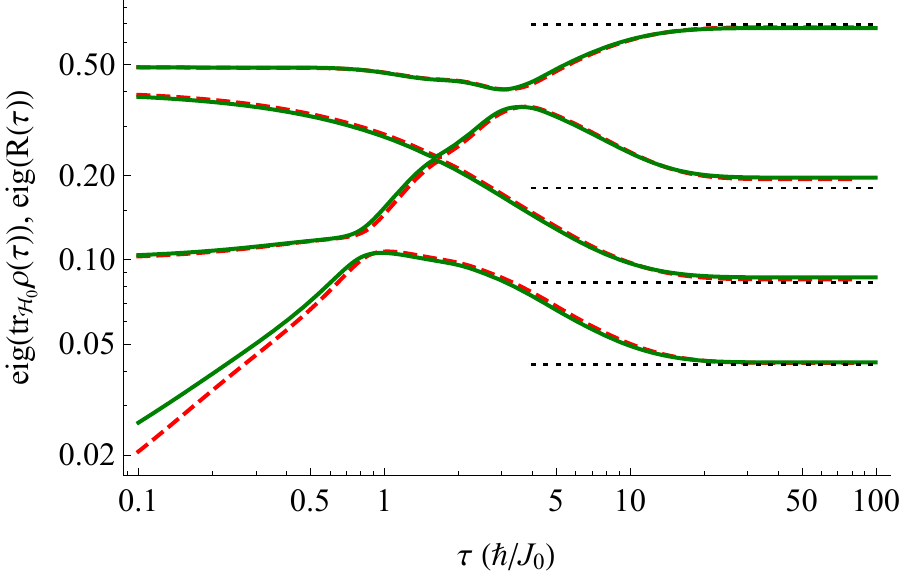}
  \end{center}
  \caption{Comparison of the LME and LME effective dynamics:
    eigenvalues of $\trzero\rho(\tau)$ (dashed lines) and of $R(\tau)$
    (solid lines) as a function of time $\tau$. Same parameters and
    same straight dotted lines as in Fig.~\ref{fig2}.  Note the
    similarity with Fig.~\ref{fig2} except for the avoided level
    crossings.}
  \label{fig3}
\end{figure}

As a direct application of our findings, we can compute the NESS in
the Zeno limit, bypassing the solution of the LME.  Denote
$R_\infty=\lim_{\Gamma\rightarrow \infty,\tau\rightarrow \infty}
R(\tau)$. From the LME (\ref{LMEforR}) we have
\begin{align} [R_\infty,h_D] &=0.  \nonumber
\end{align}
If the spectrum of the dissipation projected Hamiltonian $h_D$ is
nondegenerate, then $h_D$ and $R_\infty$ share the same set of
eigenvectors $\ket {\al}$.  It follows that
\begin{align}
  R_\infty = \sum_\al \nu_\al^\infty \ket {\al} \bra{\al}.
  \label{NESS-0}
\end{align}
Deriving from (\ref{LMEforR}) an evolution equation for the
populations of the eigenstates of $h_D$, $\nu_\al(\tau)= \bra{\al}
R(\tau) \ket{\al}$, assuming that the effective dissipator has the
canonical form $\mathcal{\tilde D}\cdot = \sum_k A_k ( {\tilde L}_k
\cdot {\tilde L}_k^\dagger -\frac{1}{2} \{ \cdot, {\tilde L}_k^\dagger
{\tilde L}_k \} )$ starting from the state $R(\tau)= \sum_\al
\nu_\al(\tau) \ket {\al} \bra{\al}$, we obtain in the Zeno limit
\begin{align}
  \frac{\partial \nu_\al(\tau) }{\partial \tau} &=\sum_{\be\neq \al}
  w_{\be \al} \nu_\be - \nu_\al \sum_{\be\neq \al} w_{\al \be},
  \label{NESS-ClassME} \\
  w_{\be \al} &= \frac{1}{\Gamma}\sum_k A_k \left| \bra{\al}
    \tilde{L}_k \ket{\be}\right|^2.
  \label{NESS-ClassME1}
\end{align}
We recognize Eq.~(\ref{NESS-ClassME}) as the classical master equation
of a Markov process with transition rates $w_{\al \be}$.  This is a
manifestation of the well-known fact that a part of the degrees of
freedom of the LME evolves in time via a classical Markov
process~\cite{Petruccione,Lesanovsky2013}. Perron-Frobenius theorem
guarantees an existence of a time-independent steady state solution of
Eq.~(\ref{NESS-ClassME}), with non-negative entries $\nu_\al^\infty$.
After normalization $\sum_\al \nu_\al^\infty=1$, the coefficients
$\nu_\al^\infty$ acquire the double meaning of eigenvalues of the
reduced NESS (\ref{NESS-0}), and steady state probabilities in the
associated classical Markov process, see Fig.~\ref{fig2} for an
illustration.  Note that by diagonalizing $h_D$ one gets both the
eigenvectors $\ket{\al}$ of $R_\infty$ and the transition rates
$w_{\al\be}$ (and, therefore, the eigenvalues $\nu_\al^\infty$).
Thus, the problem of finding the NESS, which generically requires the
diagonalization of the full Lindbladian, represented by a
non-Hermitian matrix of size $d^2 \times d^2$, reduces, in
  the Zeno limit, to the diagonalization of the Hermitian matrix
$h_D$, of size $d_1 \times d_1$ with $d_1<d$. In the example discussed
in the Sec.~\ref{sec::BoundaryDissipation}, we have $d=2^N$ and
$d_1=d/4$.

\section{Conclusions}
One might be concerned that, since our results hold in the Zeno limit,
an impractically strong dissipation must be provided. However,
one-dimensional quantum many-body systems with dissipation acting on a
few degrees of freedom are well suited for an effective Zeno
description whenever their size is sufficiently large.  To see this
fact, consider a one-dimensional system of size $N$ with local
interactions and dissipation acting near the edges. Let $\Gamma$ be
the finite strength of the dissipation.  A perturbation spreads with
finite speed (see e.g. Lieb-Robinson bound~\cite{lieb1972}), so that
the relaxation time of the system toward the global steady state
increases at least linearly with the system size,
$\tau_\mathrm{bulk}\sim N \hbar/J_0$, see, e.g., Ref.~\cite{Znidaric2015},
while the relaxation of the edges takes a time of the order
$\tau_\mathrm{diss}\sim 1/\Gamma$.  Here, $J_0$ is a factor which
fixes the energy scale associated to the Hamiltonian of the system.
For arbitrary $\Gamma$ and sufficiently large $N$, that is
\begin{align}
  \frac{\hbar\Gamma}{J_0} \gg \frac{1}{N},
  \label{ZenoRegimeValidity}
\end{align}
the system enters an effective Zeno regime $\tau_\mathrm{diss}\ll
\tau_\mathrm{bulk}$, so the NESS of the system should be well
approximated by the NESS computed in the Zeno limit $\Gamma \rightarrow
\infty$. For a few cases for which exact results are known, validity
of the (\ref{ZenoRegimeValidity}) can be demonstrated, see, e.g.
Refs.~\cite{ProsenExact2011,PhysRevE.88.062118}.  However if the Zeno NESS
is protected by extra symmetries, singular NESS behavior can happen.

\begin{acknowledgments}
  VP thanks the Department of Physics of Sapienza University of Rome
  for hospitality and financial support.  Financial support from the
  Deutsche Forschungsgemeinschaft and from the interdisciplinary UoC
  Forum ``Classical and quantum dynamics of interacting particle
  systems'' of the University of Cologne is gratefully acknowledged.
  VP and SE thank C. Kollath for discussions.

\end{acknowledgments}

\appendix

\section{Proof of Eq.~(\ref{Dyson2-Final})}
\label{SuppMat}
Here, we compute the Dyson series up to the second order of the
perturbation $K$. The calculation follows in part
Ref.~\cite{2014Venuti} and uses a similar notation.

The time-evolution of the state $\rho = \psi_0 \otimes R$ is be
defined via a Dyson series for $ \mathcal{E}(t) \mathcal{P}_0$. Up to
the second order of the Dyson series, we have
\begin{widetext}
  \begin{align}
    \label{DE}
    \mathcal{E}(t) \mathcal{P}_0 &= e^{\mathcal{L}_0 t} \left(1 +
      \int_0^t dt_1 e^{-\mathcal{L}_0 t_1} K e^{\mathcal{L}_0 t_1} +
      \int_0^t dt_1 e^{-\mathcal{L}_0 t_1} K e^{\mathcal{L}_0 t_1}
      \int_0^{t_1} dt_2 e^{-\mathcal{L}_0 t_2} K e^{\mathcal{L}_0 t_2}
      + \dots \right) \mathcal{P}_0
    \nonumber \\
    &= \mathcal{P}_0 + e^{\mathcal{L}_0 t}\int_0^t dt_1
    e^{-\mathcal{L}_0 t_1} K e^{\mathcal{L}_0 t_1}\mathcal{P}_0 +
    e^{\mathcal{L}_0 t} \int_0^t dt_1 e^{-\mathcal{L}_0 t_1} K
    e^{\mathcal{L}_0 t_1} \int_0^{t_1} dt_2 e^{-\mathcal{L}_0 t_2} K
    e^{\mathcal{L}_0 t_2} \mathcal{P}_0.
  \end{align}
\end{widetext}
In passing from the first to the second line we have used the obvious
relation
\begin{align}
  \label{DE1}
  e^{\mathcal{L}_0 t} \mathcal{P}_0 = \mathcal{P}_0 e^{\mathcal{L}_0
    t} = \mathcal{P}_0.
\end{align}
Let us focus on the second term of Eq.~(\ref{DE}) and insert the
identity decomposition $I=\mathcal{Q}_0+\mathcal{P}_0$:
\begin{align}
  \label{DE2}
  &e^{\mathcal{L}_0 t}\int_0^t dt_1 e^{-\mathcal{L}_0 t_1} K
  e^{\mathcal{L}_0 t_1}\mathcal{P}_0 \nonumber \\ &\qquad =
  e^{\mathcal{L}_0 t}\int_0^t dt_1 e^{-\mathcal{L}_0 t_1}
  (\mathcal{P}_0 + \mathcal{Q}_0) K e^{\mathcal{L}_0 t_1}\mathcal{P}_0
  \nonumber \\
  &\qquad = t \mathcal{P}_0 K \mathcal{P}_0 + e^{\mathcal{L}_0
    t}\int_0^t dt_1 e^{-\mathcal{L}_0 t_1} \mathcal{Q}_0 K
  \mathcal{P}_0.
\end{align}
In the second term of Eq.~(\ref{DE2}), we split the integral
\begin{align}
  &e^{\mathcal{L}_0 t}\int_0^t dt_1 e^{-\mathcal{L}_0 t_1}
  \mathcal{Q}_0 K \mathcal{P}_0 \nonumber \\ &\qquad =
  e^{\mathcal{L}_0 t} \left( \int_0^{{- \infty}} dt_1 \dots + \int_{{-
        \infty}}^t dt_1 \dots \right) \nonumber \\ &\qquad =
  e^{\mathcal{L}_0 t} \int_0^{{- \infty}} dt_1 e^{-\mathcal{L}_0 t_1}
  \mathcal{Q}_0 K \mathcal{P}_0 \nonumber \\ &\qquad\qquad -
  \int_t^{{- \infty}} dt_1 e^{\mathcal{L}_0 (t-t_1)} \mathcal{Q}_0 K
  \mathcal{P}_0,
\end{align}
and, after the substitutions $t_1 \to -\tilde{t}_1$, $dt_1 \to
-d\tilde{t}_1$, we obtain
\begin{align}
  \label{DE3}
  &e^{\mathcal{L}_0 t}\int_0^t dt_1 e^{-\mathcal{L}_0 t_1}
  \mathcal{Q}_0 K \mathcal{P}_0 \nonumber \\ &\qquad =
  -e^{\mathcal{L}_0 t} \int_0^{{\infty}} d\tilde{t}_1 e^{\mathcal{L}_0
    \tilde{t}_1} \mathcal{Q}_0 K \mathcal{P}_0 \nonumber \\
  &\qquad\qquad+ \int_{-t}^{{ \infty}} d\tilde{t}_1 e^{\mathcal{L}_0
    (t+\tilde{t}_1)} \mathcal{Q}_0 K \mathcal{P}_0.
\end{align}
Next, we make the change of variable $t+\tilde{t}_1 \rightarrow u$,
$d\tilde{t}_1 \rightarrow du$ in the second integral of
Eq.~(\ref{DE3}) and obtain
\begin{align}
  &e^{\mathcal{L}_0 t}\int_0^t dt_1 e^{-\mathcal{L}_0 t_1}
  \mathcal{Q}_0 K \mathcal{P}_0 \nonumber\\ &\qquad =
  -e^{\mathcal{L}_0 t} \int_0^{{\infty}} d\tilde{t}_1 e^{\mathcal{L}_0
    \tilde{t}_1} \mathcal{Q}_0 K \mathcal{P}_0 + \int_{0}^{{ \infty}}
  du e^{\mathcal{L}_0 u} \mathcal{Q}_0 K \mathcal{P}_0.
\end{align}
Renaming $\tilde{t}_1, u \rightarrow t$, we can write
\begin{align}
  \label{DE4}
  e^{ \mathcal{L}_0 t}\int_0^t dt_1 e^{-\mathcal{L}_0 t_1}
  \mathcal{Q}_0 K \mathcal{P}_0 = \left(e^{\mathcal{L}_0 t} - I
  \right) \mathcal{S} K \mathcal{P}_0,
\end{align}
where
\begin{align}
  \mathcal{S} = - \int_0^{{\infty}} dt e^{\mathcal{L}_0 t}
  \mathcal{Q}_0
\end{align}
is the pseudo-inverse of the dissipator, namely,
\begin{align}
  \mathcal{L}_0 \mathcal{S} = \mathcal{S} \mathcal{L}_0 =
  \mathcal{Q}_0.
\end{align}
The operator $\mathcal{S}$ is bounded, since the eigenvalues of
$\mathcal{L}_0$ (apart from the nondegenerate $0$ eigenvalue which is
excluded by the multiplication with $\mathcal{Q}_0$) are nonzero and
finite.  Combining Eqs.~(\ref{DE2}) and (\ref{DE4}), we conclude
\begin{align}
  \mathcal{E}(t) \mathcal{P}_0 &= \mathcal{P}_0 + t \mathcal{P}_0 K
  \mathcal{P}_0 + \left(e^{\mathcal{L}_0 t} - I \right) \mathcal{S} K
  \mathcal{P}_0 + \ldots
  \label{E(t)P0}
\end{align}
($\ldots$ denoting contributions from second and higher orders), which
retrieves the result reported in Ref.~[1].  Equation (\ref{E(t)P0}) shows,
in particular, that the leaking outside the dissipation-free subspace
for times $t> 1/\Gamma$ is of order $1/\Gamma$, namely,
\begin{align}
  \| \rho(t) - \psi_0 \otimes \traccazero \rho(t) \| &=
  O\left(\Gamma^{-1} \right).
  \label{NormLeakageOutsideDFS}
\end{align}

The evolution inside the decoherence-free subspace is given by
$\mathcal{P}_0\mathcal{E}(t) \mathcal{P}_0$.  Making use of
Eq.~(\ref{DE1}), up to the second order Dyson term we thus obtain
\begin{align}
  &\mathcal{P}_0 \mathcal{E}(t) \mathcal{P}_0 = \mathcal{P}_0 + t
  \mathcal{P}_0 K \mathcal{P}_0 + \mathcal{P}_0 e^{\mathcal{L}_0 t}
  \nonumber \\ &\qquad\times \int_0^t dt_1 e^{-\mathcal{L}_0 t_1} K
  e^{\mathcal{L}_0 t_1} \int_0^{t_1} dt_2 e^{-\mathcal{L}_0 t_2} K
  e^{\mathcal{L}_0 t_2} \mathcal{P}_0.
  \label{DysonTermsFirstOrder}
\end{align}
Now we estimate the $O(K^2)$ contribution to
$\mathcal{P}_0\mathcal{E}(t) \mathcal{P}_0$:
\begin{align}
  &\mathcal{P}_0 e^{\mathcal{L}_0 t} \int_0^t dt_1 e^{-\mathcal{L}_0
    t_1} K e^{\mathcal{L}_0 t_1} \int_0^{t_1} dt_2 e^{-\mathcal{L}_0
    t_2} K e^{\mathcal{L}_0 t_2} \mathcal{P}_0 \nonumber \\
  &\qquad= \mathcal{P}_0 \int_0^t dt_1 \int_0^{t_1} dt_2 K
  e^{\mathcal{L}_0 t_1 -\mathcal{L}_0 t_2} K \mathcal{P}_0 \nonumber
  \\ &\qquad= \mathcal{P}_0 \int_0^t dt_1 \int_0^{t_1} dt_2 K
  e^{\mathcal{L}_0 t_1 -\mathcal{L}_0 t_2} (\mathcal{P}_0 +
  \mathcal{Q}_0) K \mathcal{P}_0 \nonumber \\ &\qquad= \frac{t^2}{2}
  (\mathcal{P}_0 K \mathcal{P}_0)^2 \nonumber \\ &\qquad\qquad+
  \mathcal{P}_0 K \int_0^t dt_1 e^{\mathcal{L}_0 t_1} \int_0^{t_1}
  dt_2 e^{
    -\mathcal{L}_0 t_2} \mathcal{Q}_0 K \mathcal{P}_0 \nonumber \\
  &\qquad= \frac{t^2}{2} (\mathcal{P}_0 K \mathcal{P}_0)^2 +
  \mathcal{P}_0 K\int_0^t dt_1 \left(e^{\mathcal{L}_0 t} - I \right)
  \mathcal{S} K \mathcal{P}_0 \nonumber \\ &\qquad= \frac{t^2}{2}
  (\mathcal{P}_0 K \mathcal{P}_0)^2 -t \mathcal{P}_0 K (\mathcal{P}_0
  + \mathcal{Q}_0) \mathcal{S} K \mathcal{P}_0 \nonumber \\
  &\qquad\qquad+ \mathcal{P}_0 K\int_0^t dt_1 e^{\mathcal{L}_0 t}
  \mathcal{S} K \mathcal{P}_0.
\end{align}
Let us concentrate on the last term of the above expression. Inserting
the identity decomposition $I=\mathcal{Q}_0+\mathcal{P}_0$ and using
Eq.~(\ref{DE1}), we have
\begin{align}
  &\mathcal{P}_0 K\int_0^t dt_1 e^{\mathcal{L}_0 t} \mathcal{S} K
  \mathcal{P}_0 \nonumber \\ &\qquad= \mathcal{P}_0 K\int_0^t dt_1
  e^{\mathcal{L}_0 t} (\mathcal{P}_0 + \mathcal{Q}_0) \mathcal{S} K
  \mathcal{P}_0 \nonumber \\ &\qquad= t \mathcal{P}_0 K \mathcal{P}_0
  \mathcal{S} K \mathcal{P}_0 + \mathcal{P}_0 K\int_0^t dt_1
  e^{\mathcal{L}_0 t} \mathcal{Q}_0 \mathcal{S} K \mathcal{P}_0.
\end{align}
Gathering all terms of order $K^2$, we conclude
\begin{align}
  &\mathcal{P}_0 e^{\mathcal{L}_0 t} \int_0^t dt_1 e^{-\mathcal{L}_0
    t_1} K e^{\mathcal{L}_0 t_1} \int_0^{t_1} dt_2 e^{-\mathcal{L}_0
    t_2} K e^{\mathcal{L}_0 t_2} \mathcal{P}_0 \nonumber \\
  &\qquad= \frac{t^2}{2} (\mathcal{P}_0 K \mathcal{P}_0)^2 - t
  \mathcal{P}_0 K\mathcal{Q}_0 \mathcal{S} K \mathcal{P}_0 \nonumber
  \\ &\qquad\qquad+ \mathcal{P}_0 K\int_0^t dt_1 e^{\mathcal{L}_0 t}
  \mathcal{Q}_0 \mathcal{S} K \mathcal{P}_0.
  \label{DysonTermsSecondOrder}
\end{align}
In the last term of Eq.~(\ref{DysonTermsSecondOrder}), the integral
over time converges, thus this term is of order $\| K^2 \|=
O(1/\Gamma^2)$ and can be neglected. Bringing together
Eqs.~(\ref{DysonTermsFirstOrder}) and (\ref{DysonTermsSecondOrder}),
we obtain Eq.~(\ref{Dyson2-Final}).


%

\end{document}